
%
%
%
\input AAmac
%
%
\hyphenation{Pij-pers}
%
\newcount\eqnumber
\eqnumber=1
\def\neqn{{\rm(\the\eqnumber)}\global\advance\eqnumber by 1}
\def\refeq#1){\advance\eqnumber by -#1 {\rm(\the\eqnumber)} \advance
\eqnumber by #1}
\def\eqnam#1#2{\immediate\write1{
\xdef\ #2{(\the\eqnumber}}\xdef#1{(\the\eqnumber}}
\newcount\fignumber
\fignumber=1
\def\nfig{\global\advance\fignumber by 1}
\def\refig#1{\advance\fignumber by -#1 \the\fignumber \advance\fignumber
by #1}
\def\fignam#1#2{\immediate\write1{
\xdef\ #2{\the\fignumber}}\xdef#1{\the\fignumber}}
\def\note #1]{{\bf #1]}}
\def\draft{\headline{\bf File: \jobname\hfill DRAFT\hfill\today}}
\def\ref{\par\noindent
	\hangindent=0.7 truecm
	\hangafter=1}
%
%
\def\etal{{et al.}}

\def\cf{{cf.}}

\MAINTITLE{Short time scale monitoring of SiO sources}


\AUTHOR={ F.P. Pijpers $^1$, J.R. Pardo $^2$ and V. Bujarrabal $^2$}

\OFFPRINTS{F.P. Pijpers}

\INSTITUTE{
$^1$ Uppsala Astronomical Observatory, Box 515, S-75120 Uppsala, Sweden.

\noindent
$^2$ Centro Astron\'omico de Yebes (OAN, IGN), Apartado 148, E-19080
Guadalajara, Spain
}

\DATE{Received ; accepted}
\ABSTRACT{ We present the results of a short time scale monitoring of
SiO maser emission (v=1 J=1-0 transition) in four known strong sources.
These sources were monitored nightly for a period of about a month. The
aim of these observations is to investigate the possible presence of
variations in the maser lines on time scales of a few days to weeks,
due to sound waves propagating out from the central star. If sound waves
are responsible for the mass loss of certain cool giants, as suggested
by Pijpers and Hearn (1989) and Pijpers and Habing (1989), local
variations in density and relative velocity are expected just above
the stellar photosphere. These could give rise to variations in any
narrow spectral line formed in this region, and therefore in particular
in the SiO maser lines. Our observations indicate that variations in the
line shape (leading to relative changes in the intensity of about 20\%)
occur in the SiO emission of Mira type stars, within short time scales of
10-20 days. The main component of the profile variability is consistent
with a displacement of the velocity centroid of the dominant maser peaks,
by about 1 km s$^{-1}$ in the average.
Apparent variations in the total line flux were also found, but
could be partially due to calibration uncertainties. }


\THESAURUS{08.13.2 ; 08.03.4 ; 08.15.1 ; 02.13.3 ; 13.19.5 }

\maketitle

\titlea{Introduction}

Many if not all cool giants lose matter at a rate which can be as high
as $10^{-4} M_{\sun} {\rm yr}^{-1}$ (e.g. Knapp, 1991). Some cool giants
are large amplitude radial pulsators with periods between typically 150
and 400 d, for the Miras and up to several thousands of days for OH/IR
stars (Herman \& Habing, 1985). OH/IR stars are slightly further evolved up
the asymptotic giant branch (AGB) than Miras are. Both Miras and OH/IR
stars are emitters in several maser lines. In particular, SiO maser
emission has been observed in a large number of objects.

The precise manner in which this mass loss is driven away from these
stars is still unknown. It is clear that for those stars that show
a large amplitude radial pulsation, coupled with dust formation in the
wind, the two stage process as envisaged by Wood (1979) and Willson
and Hill (1979), and modelled in more detail by Bowen (1988), works
well. However not all stars that lose mass also have large amplitude
pulsations and form dust sufficiently near to the star. For these cool
giants the alternative mechanism proposed by Pijpers and Hearn (1989)
and applied to AGB stars by Pijpers and Habing (1989)
might well be more appropriate. In this model the mass loss is driven
by sound waves propagating out into the wind. These are generated either
by high overtone radial or non-radial pulsation or by the convection in
the surface layers of the star. As these sound waves propagate outward
in the wind they transfer momentum to the wind which accelerates it
outward.

At large distances from the star it is likely that the sound waves have
dissipated. Near the star the amplitude of the sound waves will be large
and they may well develop into shock waves before dissipating. At these
distances of only a few stellar radii from the star their effects are
therefore most likely to be detected.

There are a number of problems associated with the detection of sound
waves in the winds of stars. Their velocity amplitude will not exceed
the local sound speed by very much, if at all. This means that velocity
shifts of spectral lines will be of the order of $5\ {\rm km\ s}^{-1}$.
The density variations associated with this will be quite large but
column density variations will be much smaller, since the wave length of
the sound waves is small compared to the stellar radius. Also the
horizontal scales of the waves will only be large for low spherical
harmonics of the pulsation and therefore their effect will average
out over the stellar disc. It is therefore unlikely to find large
variations for thermally excited lines (see however Bookbinder \etal,
1989, 1993).

For maser lines and in particular maser lines originating near the
stellar surface the situation is much more favourable. The largest
amplification of the maser radiation occurs for the longest path length
in the line of sight. For a spherically symmetric wind this means that
an image of the maser source at a given velocity within the range
spanned by the maser line will show a ring rather than an extended disc.
The spatial averaging which destroys any variation will therefore be less
severe. Also, the maser intensity is expected to be in general sensitive to
the column density, and the spatial scale of the masing regions is likely
to be quite small (\cf\ Colomer \etal, 1992, Bujarrabal, 1994). Small
variations in column density can therefore have a large effect on the
intensity. Both of these effects alleviate the problems associated with
the small spatial scales of sound waves. Since SiO maser emission
originates near the stellar photosphere it affords a good opportunity
for the detection of sound waves.

SiO masers in evolved stars are known to present strong variations in
intensity (e.g. Alcolea 1993). This long-period variability essentially
follows the stellar pulsating activity, but with a poor regularity and
also presenting changes in the line shape. Variations in time scales of
days have been reported (e.g Balister et al. 1977), but not studied with
sufficient detail and sensitivity
to obtain conclusive results on their properties. The
purpose of this paper is to improve our knowledge on such rapid changes,
in particular making use of the much more powerful instrumentation
available at present. As is shown below, the data allow inferences to
be made on the possible effects of sound waves on the conditions in
the maser emitting region. Four known
strong maser sources were chosen to obtain the best possible S/N for
the line spectra~: R Leo, R Cas, Ori A, and VY CMa. Ori A is a star
forming region and is considered less likely to show any variations
on short time scales. VY CMa is an M5 supergiant star with intense SiO
emission and irregular intensity variations.
The J=1-0 v=1 $^{28}$SiO line for this star is very broad and has many
peaks. R Cas and R Leo are M6 and M7 Mira-type stars with periods of
431 and 312 days, respectively. During our
observations, the phases of these stars advanced from 0.96 to 0.09
for R Leo and from 0.14 to 0.23 for R Cas. At this epoch emission was
strong for the two stars because the SiO maximum occurs usually with a
phase delay of 0.1-0.2 (see Alcolea 1993).

\titlea{Instrumentation and observations}

We have observed the v=1 J=1-0 SiO transition (at 43 GHz) in the well
known sources R Cas, R Leo, VY CMa and Ori A.
The observations were carried out in 1992 Aug.-Sept. with the 13.7-m
radiotelescope of the Centro Astron\'omico de Yebes. The telescope was
equipped with a cooled Schottky receiver with a single-sideband  (SSB)
system temperature of 260 K. The backend was a 256x50kHz filter bank. A
polarizer was installed in front of the receiver allowing the observation
of circular polarization (the SiO masers are known to present some degree
of linear polarization).
The total line width of the emission from R Leo and R Cas is narrower
than that of VY CMa and Ori A. For this reason the latter two stars were
observed using the position switching method.
For R Leo and R Cas the frequency switching mode reducing the velocity
coverage to $\sim$ 44 km/s was used. The latter method provides spectra
with less noise.

The main beam antenna temperatures have been multiplied by 90 to get
a maser flux in Janskys.

Our monitoring began on August 22 and concluded on September 30.
Observations for each star were made typically every 24 hours, except
for R Cas which was observed twice a day. In order to improve the relative
calibration we observed each object at the same (average) elevation. This
procedure also ensures that the projection of the polarization angle
of the emission on the receiver is always the same for each source, to
avoid problems with possible partial detection of linear polarization (the
effects of the line polarization on the variability of maser emission
are discussed in detail by Mart\'{\i}nez \etal 1988). The observations of R
Cas do not satisfy this condition. This source was in general observed twice
a day, at the same elevation (50-55 degrees) but not at the same
sidereal time. We integrated on each source over about 1 or 1.5 hours.
The pointing was checked every 30 minutes by means of pseudocontinuum scans
over the maser source itself. To guarantee that the observed variations
were not due to pointing errors, most of the observations consisted in fact
of several runs of small maps of
five points in the sky~: the expected star position and four others
separated by 30". The pointing errors were then corrected every 30 minutes
by the pseudocontinuum data and, moreover, residuals were calculated
from the small maps (the corresponding very small corrections to the total
flux were taken into account a posteriori). In addition, the ambient
temperature was measured in order to check the effect of temperature
changes on the calibration. During the
period of observations the sidereal time advances 3 hours and the (average)
ambient temperature decreases by about 15 degrees.
With these data it is possible to estimate the observational errors in the
line intensity due to calibration uncertainties. This is found to be of
the order of 10$\%$.

\titlea{Data reduction}

The variation of the profiles can be expressed as a combination of a
total line flux variation and as a change of the line shape. The total
line flux variation is harder to detect because it is sensitive to
calibration errors due to e.g. variations in the calibrator temperature.
Variations in the line shape are much more straightforward to detect if
the S/N ratio per channel is sufficiently large.

Various methods are used to attempt a measurement of
intrinsic changes in the maser line flux. The height of the peak(s)
in the line profile, the area under the line and the centre of the line
profile are followed in time, as well as the ambient temperature.
The correlation of the first two with the ambient temperature is
checked in an attempt to determine apparent flux variations due to
errors in the calibration.

\begfig 14.4cm
\figure{1}{Average spectrum, peak flux and line centre of the star R Leo
during our short time scale monitoring. This star had a maximum of SiO
maser emission (v=1 J=1-0) on September 12-15, 1992 (DAY number 1 is
August 22.) }
\endfig

\begfig 14.4cm
\figure{2}{Same as Fig. 1 but for the star R Cas. This source had a
maximum of SiO maser emission (v=1 J=1-0) on August 10-15, 1992 }
\endfig

In the observations of Ori A the changes in
area and peak height are correlated with the ambient temperature
changes, with correlation coefficients between 0.7 and 0.8. The variation
observed is therefore probably for the most part due to calibration
errors rather than intrinsic variation. The line centre changed from
$6.4\ {\rm km\ s}^{-1}$ to $5.8\ {\rm km\ s}^{-1}$.

In VY CMa the correlation coefficients for area/height and ambient
temperature are between 0.8 and 0.9 and therefore again intrinsic
line flux variation cannot be determined. There is no change in
the position of the centre of the line.

In R Leo the correlation coefficients are low, between 0.5 and 0.6.
R Leo had an SiO maximum during the monitoring run, on about
september 12-15 (Fig. 1).
Since the curves for the peak height and the area as a
function of time follow the visual light curve, it is likely that
the line flux variation is mostly intrinsic. The centre of the
line profile changes from $0.4\ {\rm km\ s}^{-1}$ to $-0.1\
{\rm km\ s}^{-1}$. A similar behaviour of the line centroid has been
observed previously for R Leo (see Alcolea 1993) but with much less
detail than in the present data.

For R Cas there was no correlation of the line maxima or the area with
receiver temperature. The peak maximum and the area are both decreasing
during the monitoring period (Fig. 2),
which is probably intrinsic since the star was also becoming fainter
visually (see the Introduction for the phase of
R Cas in the observations). The position of the line centre did not
change significantly during the monitoring.

Neither in the height of the maxima of the line profile, nor in the area
could any periodic change be detected for any of the sources.

\begfig 14.4cm
\figure{3}{Residual spectra (see text) of the star R Leo corresponding to
three observations of our monitoring. Line shape variations are evident}
\endfig

To detect variations in the line profiles a mean profile is constructed
by adding all the separate line profiles for the observations and
then dividing by the number of separate observations. Subsequently this
mean profile is subtracted from each of the individual spectra after being
scaled so that it always has the same total line flux as the individual
spectrum. Using channels far outside of the maser line as a reference for
the noise per channel, a $\chi^2$ test is then performed on the residual
amplitude within the channels designated as containing the maser line
emission. These channels are chosen by inspection of the mean profile.
The $\chi^2$ tests confirm unambiguously that the variations found with
respect to the average profile are statistically significant for R Leo and
R Cas. In VY CMa no line shape variations were detected. In Ori A a
significant variability was found, but will not be discussed here because
of the different nature of the source.
The average spectrum, and the variations found in the peak flux and
profile centroid are shown in Figs. 1 and 2 for the stars R Leo and
R Cas, respectively. Some selected residual profiles are shown (Fig. 3).
Even without the results
of the $\chi^2$ test it is clear that the line profiles change on a
time scale of days. The residual line profile after subtraction of the
mean is well above the noise on many separate observations.
Plots of the time variation of individual channels are shown in Fig. 4
for R Leo, the star for which the best results were obtained. In some
channels a sinusoidal variation around the mean might be
present, indicating a period of $\geq 30\ {\rm d}$. However the total
time span of this monitoring is too short to confirm this and any
periodicity in the signal is therefore only tentative. On the other
hand, it should be noted that the detection of a variability with longer
periods is probably difficult (unless it is very clear), since the well
known and intense long-period variations will contaminate the observation.
The SiO maser in R Cas shows a similar behavior to that in R Leo (Fig. 5),
but the effect is no clearer.
The variations are smaller than in R Leo and the observational noise is
relatively more important. The variations found in individual channels
for these two sources range between 10 and 30 \% of the peak flux, and the
typical variation time is of about 10-20 d. It is
remarkable in the data shown in Figs. 4 and 5 that the maximal variations
are found around the two main peaks of R Leo and
the main peak of R Cas, and with opposite sign from one side of the
centroid to the other. Such a behaviour could be understood as due to a
global displacement of the peaks, by 1 km s$^{-1}$ in the average. Note
that the relative amplitude of the the detected variations is small in
all cases, and the intensity and overall structure of the profiles differ
only slightly from the averages shown in Figs. 1 and 2.

A very short time scale monitoring of the star R Cas (on a time scale
of hours) was also performed. No clear results could be extracted from
these data. Some variations of the intensity were observed but these
were correlated with the ambient temperature and source elevation.
Intrinsic variations on time scales of hours were not detectable because
of the observational uncertainties.

Summarizing; for VY CMa no line profile variations could be
detected. The channel noise is too large. Both R Cas and R Leo do show clear
line profile variations on a time scale of days. If there is any period
in the data it is likely to be in the range of 30-60 days. The total
time span of this monitoring program was too short to make any definite
conclusions.

\titlea{Regarding the origin of the short time scale variability of SiO
masers}

\begfigwid 12.0cm
\figure{4}{Variation of the difference between individual spectra and
the average for the star R Leo. The time evolution is shown of individual
channels within the velocity range of the line emission.
The strongest variations are found near the peaks, the main peak in channel
59 (1) and a secondary in channel 73 (2) (DAY number 1 is
August 22.) }
\endfig

\begfigwid 12.0cm
\figure{5}{Same as Fig. 4 but for the star R Cas. Note that the variations
are smaller than in R Leo and there is even less evidence for periodic
behavior. The three main peaks of the average spectrum are centred at
channels 72, 84 and 60 }
\endfig

The cause of the variations reported here could be simply the underlying
light curve of the star which changes the input for the maser
radiation. This however is more likely to produce a change in total
maser line flux than in the line shape if the spatial structure of the
masing region is really uniform and constant.
Also, the period found should then be the same as the visual
period of the star. Such long time scale changes in the total line flux
were indeed detected in R Cas and probably in R Leo as well.

If the masing region is not uniform then differences in light travel
time for different masing patches with differing radial velocities,
coupled with a rapidly varying central source of maser pumping photons,
could cause line profile variations. This is similar to what is
seen in the broad emission lines of active galactic nuclei (AGN)
If this were the case one might attempt the type of reverberation
mapping that was developed for novae (Coudrec, 1939) and has been used
for AGN (Peterson, 1993) to deduce the velocity field of the gas of
which the masing patches are tracers. However for a time-scale of
variation of days the difference in path length for different patches
would be a few hundred stellar radii. This is far larger than the size
of the SiO masing region deduced from mapping which is less than one
light hour.

If sound waves are present in the wind and if they are generated by
convection they are the equivalent of the solar 5 min oscillations.
The period of oscillations always scales inversely proportional to the
square root of the mean density because of the period-mean density
relation (see e.g. Cox, 1980)~:
$$
\Pi\sqrt{\overline{\rho}/\overline{\rho}_{\sun}}\ =\ Q
\eqno\neqn$$
The density $\overline{\rho}$ refers to the mean density of the star.
The mass of cool giant stars is not more than a few solar masses and
the radius is several hundred solar radii. Radius estimates from lunar
occultation measurements are available for R Leo (Di Giacomo \etal, 1991),
which yields $380\ R_\odot$. Photometric analysis (e.g. Cahn \& Elitzur,
1979) yields a similar value for R Cas. A mass of $1.5\ M_\odot$ does
not seem unreasonable for these stars (\cf\ Wood, 1974~; Bowen, 1988~;
Boothroyd \& Sackmann, 1988). The equivalents of the solar 5 min
oscillations should then have a period of $\sim 21\ {\rm d}$. However
the relative depth of the convective layer in cool giants is larger
than it is in the sun. The current best estimates from helioseismology
(\cf\ Guzik \& Cox, 1993, Roxburgh \& Vorontsov, 1993) give a radius of
the bottom of the solar convective zone between $R = 0.71 - 0.72\ R_{\sun}$.
The models of Wood (1974) put the bottom of the convective layer in red
giants close to $R = 0.2\ R_*$. As Schwarzschild (1975) argues, the
peak of the energy spectrum of sound waves generated by turbulent
convection may well shift to longer wave lengths because of a scaling
with the total relative depth of the convective layer. This means that
the most strongly excited sound waves propagating into the atmosphere
and the wind may have a larger wave length and therefore a larger period
by roughly a factor of $3$ that such a simple scaling with the size of
the convective layer suggests. The range of possible periods is therefore
$\sim 21-63\ {\rm d}$.

Given an atmospheric sound speed of typically $5\ {\rm km/s}$ the wave
length of sound waves with a period in this range is $\sim 9\ 10^9 -
3\ 10^{10}\ {\rm m}$. The horizontal wave length of the wave fronts
can be larger than the vertical wave length. This depends on the value of
the indices $l$ and $m$ of the spherical harmonics of the non-radial
oscillations that generate the sound waves propagating into the wind.
If for simplicity it is assumed that the two size scales are the same
then the number of wave crests and valleys over a stellar disc is
$750 - 5000$. In the continuum or in a thermal line any effect due to
sound waves is therefore at best only detectable at $\mu$mag-level
accurate photometry (\cf\ Belmonte \etal, 1990). The advantage of using
maser lines is that the masing effects tend to amplify the differences
in the path length. If one constructs a map of a source with a shell
structure the overal angular distribution of brightness is a ring rather
than a disc. Circumstellar SiO masers are thought to be such shell
sources (Bujarrabal, 1994). The spatial averaging for a sound wave front
passing through this ring is over a number of wave crests and valleys
(or `elements') proportional to the $\ell$ of the spherical harmonic.
Using the same assumption of equal horizontal and vertical size scales,
the number of elements in the ring of maser emission is $60 - 200$.
In each of these elements the amplitude of the density and velocity
variations are related according to~:
$$
{\delta \rho\over\rho}\ =\ {\delta v\over c_S}
\eqno\neqn$$
where $c_S$ is the sound speed. During their passage away from the star
the sound waves will deform into shock waves. The limiting amplitude of
these shock waves is determined by two opposing factors. One is the
decrease in background density which will act to increase the wave
amplitude which can be seen from arguments of conservation of energy.
The other is radiative damping which tends to decrease the amplitude.
The combination of these effects is likely to produce wave amplitudes
of order unity (see e.g. Koninx \& Pijpers, 1992) in the region of
interest. Due to non-linear effects the change in density at maximum
compression will be a factor somewhat larger than unity.
$$
{\delta \rho\over\rho}\ \geq\ {\delta v\over c_S}\ \sim\ 1
\eqno\neqn$$

The apparent size of the emitting spots of SiO masers ranges from scales
comparable to the sound wave length quoted above to about 50 times such
values (Colomer et al., 1992, Bujarrabal, 1994). The maser emission of the
most compact spots is known to amount to about one half of the total flux
and it consists of spectral spikes with typical widths of about
$1\ {\rm km\ s}^{-1}$. This emission is very probably saturated in which case
it is strongly dependent on the contribution of its "unsaturated core"
inside the amplifying path.
Since the size of each of such a cores is thought not to be much larger
than the sound wave length, the intensity and velocity of the spectral
features emitted by these compact spots should vary significantly with
the passage of sound waves.
The effects of the waves on the intensity of the extended (less saturated)
emission is not clear. Although many elements of density variation are
expected to be present in an emitting volume, even small variations in the
averaged column density can strongly affect the intensity due to the
non-linear maser effect. In the unsaturated SiO masers this effect may
lead to amplification factors larger than 10$^6$.
In summary, sound waves are likely to produce relatively strong local
variations of the density and the velocity in the inner circumstellar
layers. Such phenomena should yield a significant variability in the
maser features, up to percentages in intensity not much smaller than
one half of the total. Significant variations in the velocity of the
spectral features are also expected.

Finally it should be mentioned that
the presence of sound waves in the wind does not necessarily produce
periodic variations for two reasons. The first is that there is likely
to be a broad-peaked spectrum for the wave lengths of the sound waves
just as there is in the sun. This will produce a multiperiodic signal
that cannot easily be decomposed. Furthermore, inhomogeneity of the
masing region will tend to destroy any clear periodicity in the total line
flux variation.

\titlea{Conclusions}

A short time scale monitoring is presented of SiO maser emission (v=1
J=1-0 transition) in the evolved stars R Cas, R Leo and
VY CMa, and in the star forming region Ori A.  Observations were typically
carried out every day during 40 days. Special attention is paid to the
minimization and correction of the calibration uncertainties.
The measurement of (small) variations in the total intensity is
always difficult, due to such uncertainties, however we believe that
variations in the total emission of about 10-20\% were probably detected
during our observing program. On the other hand, the variations of the line
shape are much more easy to determine. Significant line shape variations are
found on time scales of 10-20 d close to the line peaks of R
Leo and R Cas, leading to relative variations of the order of 20\% in
individual channels of the spectral detector. Such relative changes in
intensity may be quite different inside the line spectra, in fact the sense
of the variations tends to change from one side of the peaks to the other.
It is argued that such relative variations are probably not due to
changes in the stellar emission (i.e. in the maser pumping), but to changes
of the column density and/or dynamics in the emitting region.
In particular, the measured variability is consistent with global
displacements of the line peak in velocity of on the average $\sim 1
\ {\rm km}\ {\rm s}^{-1}$. Accordingly the conclusion must be that, at
least in some cool giant stars, the gas velocity and probably also the
density of the SiO emitting region changes on time scales significantly
shorter than the visual period~; of the order of 10-20 days. It is
suggested that this variation is due to the presence of sound waves in
the inner shells of circumstellar material surrounding these stars.
To better determine whether there is any periodicity present in the
short time scale line profile variations, a monitoring program with a
similar frequency of observations but much longer duration would be needed.
However, the measurement of periods larger than one month could be
severely contaminated by the variability of the masers following the
stellar cycle; a variability that is known to be often relatively irregular.

\acknow{While this research was in progress FPP was employed at the
Astronomy Unit, School of Mathematical Sciences, Queen Mary and Westfield
College, London, UK. FPP gratefully acknowledges the financial support
of the CAY for his visit and the use of its facilities for this work. This
work has been partially supported by the spanish CICYT, under project
PB90-408. }

\begref{References}

\ref
Alcolea J., 1993,
{Tesis doctoral. Universidad Aut\'onoma de Madrid.}
\ref
Balister M., Batchelor R.A., Haynes R.F., et al., 1977,
{MNRAS}
180, 415
\ref
Belmonte, J.A., P\'erez Hern\'andez, F., Roca Cort\'es, T., 1990,
{A \& A}
231, 383
\ref
Bookbinder J., Brugel E., Brown A., 1989,
{ApJ}
342, 516
\ref
Bookbinder J., Brugel E., Brown A., 1993,
preprint
\ref
Boothroyd A.I., Sackmann I.-J., 1988,
{ApJ}
328, 641
\ref
Bowen G.H., 1988,
{ApJ}
329, 299
\ref
Bujarrabal V., 1994,
{A \& A}, in press
\ref
Cahn J.H., Elitzur M., 1979,
{ApJ}
231, 124
\ref
Colomer F., Graham D.A., Krichbaum T.P., R\"onn\"ang B.O., de~Vicente P.,
Witzel A., Barcia A., Baudry A., Booth R.S., G\'omez-Gonz\'alez J.,
Alcolea J., Daigne G., 1992,
{A \& A}
254, L17
\ref
Coudrec P., 1939,
{Ann. d'Astroph.}
2, 271
\ref
Cox J.P., 1980,
{Theory of stellar pulsation},
Princeton Univ. Press, Princeton,
10
\ref
Di Giacomo A., Richichi A., Lisi F., Calamai G., 1991,
{A \& A}
249, 397
\ref
Guzik J.A., Cox A.N., 1993,
{ApJ}
411, 394
\ref
Knapp G.R., 1991,
{ASP Conf. Ser. Vol. 20 : `Frontiers of Stellar Evolution'},
p. 229
\ref
Koninx J.P.M., Pijpers F.P., 1992,
{A \& A}
265, 183
\ref
Herman H., Habing H.J., 1985,
{Physics Reports}
124, No.4, p.255
\ref
Mart\'{\i}nez A., Bujarrabal V., Alcolea J., 1988,
{A \& A Suppl. Ser.}
74, 273-298
\ref
Peterson B.M., 1993,
{PASP}
105, 247
\ref
Pijpers F.P., Habing H.J., 1989,
{A \& A}
215, 334
\ref
Pijpers F.P., Hearn A.G., 1989,
{A \& A}
209, 198
\ref
Roxburgh I.W., Vorontsov S.V., 1993,
{ASP Conf. Ser. Vol.~42~: GONG 1992 Seismic investigation of the sun and
stars},
169
\ref
Schwarzschild M., 1975,
{ApJ}
195, 137
\ref
Willson L.A., Hill S.J., 1979,
{ApJ}
228, 854
\ref
Wood P.R., 1974,
{ApJ}
190, 609
\ref
Wood P.R., 1979,
{ApJ}
227, 220
\endref

\bye